\documentclass[12pt]{article}
\usepackage{amssymb,latexsym,epsfig,cite}

\newcommand{\be}{\begin{equation}}
\newcommand{\ee}{\end{equation}}
\def\bea{\begin{eqnarray}}
\def\eea{\end{eqnarray}}
\begin{document}

\thispagestyle{empty}

\begin{center}
{\Large \bf Gamow vectors and Supersymmetric Quantum Mechanics}
\end{center}

\vskip1cm

\begin{center}
Oscar Rosas-Ortiz\\[2ex]
{\footnotesize \it Departamento de F\'\i sica, CINVESTAV-IPN, A.P.
14-740, 07000 M\'exico D.F., Mexico}
\end{center}


\begin{abstract}
Gamow solutions are used to transform self--adjoint energy operators by
means of factorization (supersymmetric) techniques. The transformed
non--hermitian operators admit a discrete real spectrum which is
occasionally extended by a single complex eigenvalue associated to
normalized eigensolutions. These new Hamiltonians are not
pseudo--hermitian operators and also differ from those obtained by means
of complex--scaling transformations. As an example, Coulomb--like
potentials are studied.
\end{abstract}

\section{Introduction}

Complex energies were studied for the first time in a paper of Gamow
concerning the alpha decay (1928) \cite{Gam28}. In a simple picture, a
given nucleus is composed in part by alpha particles (${}_2^4 He$ nuclei)
which interact with the rest of the nucleus via an attractive well
(obeying the presence of nuclear forces) plus a potential barrier (due, in
part, to repulsive electrostatic forces). The former interaction
constrains the particles to be bounded while the second holds them inside
the nucleus. The alpha particles have a small (non--zero) probability of
tunneling to the other side of the barrier instead of remaining confined
to the interior of the well. Outside the potential region, they have a
finite lifetime. Thus, alpha particles in a nucleus should be represented
by {\it quasi--stationary\/} states. For such states, if at time $t=0$ the
probability of finding the particle inside the well is unity, in
subsequent moments the probability will be a slowly decreasing function of
time (see e.g. Sections~7 and 8 of reference \cite{Fock}).

In his paper of 1928, Gamow studied the escape of alpha particles from the
nucleus via the tunnel effect. In order to describe eigenfunctions with
exponentially decaying time evolution, Gamow introduced energy
eigenfunctions $\psi_G$ belonging to complex eigenvalues $Z_G = E_G -
i\, \Gamma_G$, $\Gamma_G>0$. The real part of the eigenvalue was
identified with the energy of the system and the imaginary part was
associated with the inverse of the lifetime. Such `decaying states' were
the first application of quantum theory to nuclear physics.

Three years later, in 1931, Fock showed that the law of decay of a
quasi--stationary state depends only on the energy distribution function
$\omega(E)$ which, in turn, is meromorphic \cite{Fock}. According to Fock,
the analytical expression of $\omega(E)$ is rather simple and has only two
poles $E= E_0 \pm i \, \Gamma$, $\Gamma >0$ (see equation (8.13) of
reference \cite{Fock}). A close result was derived by Breit and Wigner in
1936. They studied the cross section of slow neutrons and found that the
related energy distribution reaches its maximum at $E_R$ with a
half--maximum width $\Gamma_R$. A resonance is supposed to take place at
$E_R$ and to have ``half--value breath'' $\Gamma_R$ \cite{Bre36}. The
resonances can be defined as eigensolutions $\psi_R$ of the Hamiltonian
with complex eigenvalue $z_R = E_R - i\, \Gamma_R/2$. This complex number
also corresponds to a first--order pole of the $S$ matrix \cite{Hei47}
(for more details see e.g. \cite{del03}). However, as the Hamiltonian is a
Hermitian operator, then (in the Hilbert space ${\cal H}$) there can be no
eigenstate having a strict complex exponential dependence on time. In
other words, decaying states are an approximation within the conventional
quantum mechanics framework. This fact is usually taken to motivate the
study of the rigged (equipped) Hilbert space $\bar {\cal H}$ \cite{Boh89}.
The mathematical structure of $\bar {\cal H}$ lies on the nuclear spectral
theorem introduced by Dirac in a heuristic form \cite{Dir58} and studied
in formal rigor by Maurin \cite{Mau68} and Gelfand and Vilenkin
\cite{Gel68}.

Some other approaches extend the framework of quantum theory so that
quasi--stationary states can be defined in a precise form. For example,
the complex--scaling method \cite{Agu71,Sim72,Gir03} (see also
\cite{Sud78}) embraces the transformation $H \rightarrow SHS^{-1} =
H_{\theta}$, where $S$ is the complex--scaling operator $S=e^{-\theta
rp}$, $[r,p]=i$, such that $Sf(r) = f(r e^{i\, \theta})$. This
transformation converts the description of resonances by non--integrable
Gamow states into one by square integrable states (A relevant aspect of
the method is that it is possible to construct a resolution to the
identity \cite{Ber73}).  Thus, the complex--scaled resonance
eigenfunctions are $\theta$--dependent so they can be normalized. 
Moreover, as the complex eigenvalues are $\theta$--independent, the
resonance phenomenon is just associated with the discrete part of the
complex--scaled Hamiltonian \cite{Moi98} (but see \cite{Sud78}).

In this paper we show that Gamow (decaying) eigensolutions can be used to
transform Hermitian Hamiltonians into non--self adjoint energy operators
with purely real spectrum or admitting a single extra complex eigenvalue
with square--integrable wavefunction. The new Hamiltonians could be
profitable as testing operators in diverse approaches including
complex--scaling and pseudo--hermitian \cite{Mos03} transformations. As we
shall see, it is not necessary to work in a equipped Hilbert space
framework because the Gamow solutions will be used merely as mathematical
tools. Moreover, the exponential growing of the Gamow solutions for
large distances will be primordial in order to get well-behaved complex
potentials. The mechanism we are going to use is the factorization method
in a `complex' version \cite{Ros03}. As usual, the procedure and results
can be interpreted in terms of supersymmetric quantum mechanics.

The next section introduces general expressions for transforming
spherically symmetric potentials in terms of appropriate Gamow vectors. It
is shown that new complex potentials are derived so that their discrete
spectrum is real. The Coulomb potential is managed as example. The
Section~3 shows how the approach can be generalized to include an extra
single complex eigenvalue into the initial discrete spectrum. The related
eigensolution is then shown to be of finite norm. Finally, Section~4 is
devoted to the concluding remarks.

\section{Supersymmetric Gamow transformations}

\subsection{The complex factorization}

Let us consider the time--independent Schr\"odinger equation for a
spherically symmetric potential $V(r)$. After separation of
angular variables, the equation reduces to a differential equation
involving only the radial variable:
\be
H_{\ell} \, \psi(r,\ell) = E \psi(r,\ell),
\label{eigen}
\ee
which can always be integrated numerically. The reduced
Hamiltonian reads
\be
H_{\ell} \equiv - \frac{d^2}{dr^2} + V_{\ell}(r) =
-\frac{d^2}{dr^2} + \frac{\ell(\ell +1)}{r^2} + V(r),
\label{hamil1}
\ee
where the effective potential $V_{\ell}(r)$ has the domain $D_V
=[0, +\infty)$ and the units of energy and coordinates have been
properly chosen.

The nature of the energy spectrum of $H_{\ell}$ may be deduced from the
asymptotic behaviour of the solutions $\psi(r,\ell)$ which are regular at
the origin. If $V(r)$ approaches zero asymptotically faster than $1/r$:
$\lim_{r \rightarrow \infty} rV(r) =0$, then the energy spectrum contains
two parts: (a) Negative discrete values $E_1(\ell), E_2(\ell), \ldots$ To
each of them corresponds a radial wavefunction of finite norm. (b) Unbound
continuous positive spectrum, with solutions regular at the origin but
indefinitely oscillating in the asymptotic region. On the other hand, if
$V(r)$ approaches zero as $1/r$ when $r \rightarrow \infty$, the essential
result concerning the nature of the spectrum persists \cite{Messiah}. We
shall concentrate on the discrete spectrum by assuming that a complete set
of normalized wavefunctions $\psi_n(r,\ell) \in {\cal H}$ has been given
for each $V(r)$, otherwise $H_{\ell}$ would not be an observable.

We look for a complex--type factorization \cite{Ros03} of the Hamiltonian
(\ref{hamil1}):
\be
H_{\ell} = AB + \epsilon
\label{factor1}
\ee
with factorization constant $\mathbb C \ni \epsilon =
\epsilon_1 + i \, \epsilon_2$; $\epsilon_1, \epsilon_2 \neq 0 \in
\mathbb R$ and a couple of not mutually adjoint first order
operators
\be
A:= -\frac{d}{dr} + \beta, \qquad B:= \frac{d}{dr} + \beta
\label{factors}
\ee
where $\beta$ is a complex--valued function fulfilling the Riccati
equation
\be
-\beta'(r) + \beta^2(r) + \epsilon = V_{\ell}(r).
\label{rica1}
\ee
This last equation is easily solved by means of the logarithmic
transformation $\beta(r) = -\frac{d}{dr} \ln u(r)$, with $u(r)$
the eigensolution of $H_{\ell}$ belonging to the complex
eigenvalue $\epsilon \equiv -k^2$, $\mathbb C \ni k = k_1 + i\,
k_2$; $k_1,k_2 \in \mathbb R$.

Remark that $H^{\dagger}_{\ell} = B^{\dagger} A^{\dagger} + \bar \epsilon
= H_{\ell}$ (the bar stands for complex conjugation) because the
Hamiltonian is assumed to be self-adjoint in the Hilbert space ${\cal H}$.
A relevant aspect of the complex factorization
(\ref{factor1})-(\ref{rica1}) is that the reverse ordering of the factors
gives rise to non--hermitian second order differential operators:
\be
BA + \epsilon = H_{\ell} + 2 \beta'(r) := h_{\ell}.
\label{factor2}
\ee
Conventional factorizations assume {\it a priori} $A= B^{\dagger}$ and
real $\epsilon$ (see e.g. \cite{Hul51}). In counterdistinction, complex
factorization is more in the spirit of the `refined factorizations'
reported recently \cite{refined} (see also \cite{rmf1}). The following
intertwining relationships hold
\be
h_{\ell} B = B H_{\ell}, \qquad H_{\ell} A = A h_{\ell}
\label{intertwin}
\ee
which permit to determine the solutions $\Psi \propto B \varphi$ of
$h_{\ell} \, \Psi = \lambda \Psi$, $\lambda \in \mathbb C$, by giving the
solutions $\varphi$ of $H_{\ell} \, \varphi = \lambda \varphi$. The
operator $A$ reverses the action of $B$. In the supersymmetric language,
$H_{\ell}$ and $h_{\ell}$ are understood as supersymmetric partners while
$\beta(r)$ is the superpotential (see e.g. \cite{And84} and references
quoted therein).

In general, we want to keep the physical interpretation of $\Psi$ as
connected with the probability density $\rho(r) = \vert \Psi (r) \vert^2$
in ${\cal H}$ (The dependence of $\Psi$ on $\ell$ will be always
implicitly considered). Hence, we look for functions
\be
\Psi \propto B \varphi = \frac{W(u, \varphi)}{u}
\label{newsol}
\ee
which are square--integrable in ${\cal H}$ (the symbol $W(\cdot,\cdot)$
stands for the wronskian of the involved functions). Of course, this last
condition is not imperative in equation (\ref{newsol}). For instance, one
could extend the initial boundedness condition $\vert \psi(r,\ell)\vert^2
< \infty$ to better admit another kind of normalization in order to
generalize selfadjointness (e.g., in the picture of a equipped Hilbert
space $\bar {\cal H}$). But, in this way, the physical interpretation of
either $\psi(r,\ell)$ or $\Psi(r)$ as wavefunctions is less clear (one
dimension plane waves, for example, are known to be not in $L^2(\mathbb
R)$ but having a probability density which is everywhere finite in the
Dirac sense. In other words, the plane waves could be understood as energy
Dirac vectors in $\bar {\cal H}$. However, if we apply realistic vanishing
boundary conditions at $x=0$ and $x=L$, or $L$--periodic boundary
conditions, the plane waves can be normalized in the conventional form.
Thus, `free particles' are but an abstraction from the actual quantum
world).

As it could be expected, the set of eigenvectors (\ref{newsol}) is
uncommon in ${\cal H}$: though they can be normalized, their elements are
not mutually orthogonal \cite{Ros03} (An optional bi--orthogonal basis has
been recently discussed in \cite{Mun03}). These vectors are natural in the
spaces with an indefinite metric as studied in the Pontrjagin--Krein
formalism \cite{Goh69} (see also \cite{Ram03}).

\subsection{Gamow transformations}

Let us show how the Gamow solutions can be used as transformation
functions $u(r)$ in equation ({\ref{newsol}). First, following
Gadella--de~la~Madrid, we define a Gamow function as a solution of the
time--independent Schr\"odinger equation with complex eigenvalue and
purely outgoing boundary conditions \cite{Mad02}. Thus, if $u(r)$ is such
that $u(r=0)=0$, $u(r \rightarrow +\infty) \sim e^{-kr}$ $(k_1 <0)$, and
solves (\ref{eigen}), (\ref{hamil1}) with $E=\epsilon \in {\mathbb C}$,
then $u(r)$ is a Gamow solution (Observe that $\epsilon$ does not
necessarily correspond to the poles of the $S$ matrix!). In the context of
the alpha decay, the condition $u(r=0)=0$ describes the `creation' of
alpha particles inside the nucleus and obeys the fact that there cannot be
any transmission into the region $r<0$ because the effective potential is
infinite there (i.e., this condition avoids the incoming probabilities and
is related with the adjointness of the Hamiltonian \cite{Mad02}). On the
other hand, the outgoing boundary condition ensures the decay rate obeyed
by the particles after tunneling the electrostatic barrier.

Let us take ${\rm Re}(\epsilon) \equiv E_R = k_2^2 -k_1^2 > 0$ in
$\epsilon = (k_2^2 -k_1^2) - 2 i \, k_1 k_2$. Thus $\vert k_2 \vert >
\vert k_1 \vert$. We can distinguish two general cases:

\begin{enumerate}
\item[1)] 
$k_1 <0, k_2 <0$. Here $\epsilon^- = E_R - i\, \Gamma^-/2$, with $\Gamma^-
= 4 k_1 k_2 >0$, is associated with the decaying part of the solution
$U(t) \vert \phi_{\epsilon^-} \rangle = e^{-i\, t \, E_R} \, e^{-t \,
\Gamma^- /2} \vert \phi_{\epsilon^-} \rangle$.

\item[2)] 
$k_1 <0$ and $k_2 >0$. The complex energy $\epsilon^+ = E_R + i\,
\Gamma^+/2$, with $\Gamma^+ = 4 \vert k_1 \vert \, k_2 >0$, is associated
with the growing part of the solution $U(t) \vert \phi_{\epsilon^+}
\rangle = e^{-i\, t \, E_R} \, e^{t\, \Gamma^+ /2} \vert \phi_{\epsilon^+}
\rangle$.

\end{enumerate}

In both cases the roles are interchanged under complex conjugation. Now,
if $\epsilon^{\pm}$ correspond to the poles $z_R^{\pm}$ of the S matrix,
then the lifetime $\tau = 1/\Gamma^-_R$ decreases as the energy increases.
Thus, for small widths (large lifetime) the energy resonances are close to
the real axis and the Gamow vectors could be considered as bounded states
for certain physical phenomena. On the other hand, as $\Gamma^-_R$
increases, the resonances move away from the real axis and the Gamow
vectors are far to be considered as representative of bound states.

Now, let us analyze in detail the equation (\ref{newsol}). Our goal is to
characterize the spectrum of $h_{\ell}$ as well as its eigenfunctions in
terms of the analytical behaviour of $\varphi (r, \ell)$ and the boundary
conditions of $u(r)$. 

A direct calculation shows that $u(r) \propto r^{\ell +1}$ satisfies
$u(r=0) =0$. Thereby, equation (\ref{newsol}) reads
\be
\Psi(r<<1) \sim \varphi'(r<<1) - \frac{\ell +1}{r} \,
\varphi(r<<1).
\label{cota2}
\ee
It is clear that $\Psi(r)$ will be regular at the origin if $\varphi$ is
such that $\varphi (r<<1) \sim r^s$, $s \geq 1$. In other words, if
$\varphi$ is regular at the origin then $\Psi(r=0)=0$.

The purely outgoing boundary condition, in turn, is equivalent to the
following expression (see \cite{Mad02} p 630):
\be
\lim_{r \rightarrow \infty} \frac{d}{dr} \ln u(r) = - \lim_{r
\rightarrow \infty} \beta(r) = - k.
\label{out}
\ee
Hence, equation (\ref{newsol}) reduces to
\be
\lim_{r \rightarrow \infty} \Psi \propto \lim_{r \rightarrow
\infty} \varphi' + k \, \lim_{r \rightarrow \infty} \varphi.
\label{cota1}
\ee
As the solution $\varphi$ grows indefinitely as one of either $e^{\pm
\kappa r}$, $\kappa = \sqrt{- \lambda}$, we can identify the following
cases:

\begin{enumerate}
\item[I)] 
For a (denumerably infinite) set of negative discrete values $\lambda \in
\{ E_n(\ell) \}$, the solution $\varphi$ in (\ref{cota1}) behaves as
$\varphi \sim e^{-\kappa r}$, $\kappa >0$. Thus $\Psi \sim (k - \kappa)
e^{-\kappa r}$, $\kappa>0$.

\item[II)] 
If $\lambda >0$, then $\varphi \sim \sin (\kappa r - \frac{\ell \pi}{2} +
\delta_{\ell})$, with $\delta_{\ell}$ the phase shift. Thus, $\varphi$ is
an acceptable eigensolution of $H_{\ell}$ for any $\lambda >0$ and
represents an unbound state \cite{Messiah}. Hence, if $\lambda >0$ then
$\Psi(r)$ indefinitely oscillates when $r \rightarrow \infty$.

\item[III)]
If $\lambda \in \mathbb C$ then equation (\ref{cota1}) gives $\Psi_{\pm}
\sim (\pm \kappa + k) e^{\pm \kappa r}$. Moreover, if $\lambda = \epsilon$
(equivalently $\kappa = k$) then $\Psi_- =0$ and $\Psi_+ \sim 2ke^{kr}$.
The former solution is rather trivial as $W(u,u)=0$ in equation
(\ref{newsol}). Now, as $k_1 <0$, it seems that $\Psi_+$ could satisfy
$\lim_{r \rightarrow +\infty} \vert \Psi_+ \vert =0$. However, in such a
case, $\varphi_+$ should also satisfy both conditions $\varphi_+(0)=0$ and
$\varphi_+ \propto e^{kr}$, $k_1<0$, which is not possible since $\lambda$
is complex and $H_{\ell}$ is a selfadjoint operator in $H_{\ell} \varphi_+
= \lambda \varphi_+$. A similar situation arises for any complex number
$\lambda$ different from $\epsilon$.

\end{enumerate}

In summary, for Gamow transformation functions in (\ref{newsol}), if
$\varphi \in L^2(\mathbb R^+)$ then $\Psi \in L^2(\mathbb R^+)$.
Furthermore, equation (\ref{newsol}) does not produce eigenfunctions of
the non--hermitian Hamiltonian $h_{\ell}$ belonging to complex
eigenvalues. Thereby, the complete discrete spectrum $\sigma_d(H_{\ell})$
of the initial Hamiltonian $H_{\ell}$ is inherited to the Gamow
transformed Hamiltonian $h_{\ell}$. In order to exhaust our analysis, let
us consider the complex factorization (\ref{factor2}). It is easy to
verify that the kernel of $A$ provides an eigenfunction
$\xi_{\epsilon}(r)$ of $h_{\ell}$ belonging to $\epsilon \in \mathbb C$.
Thus, $\xi_{\epsilon} \propto 1/u$ fulfills $h_{\ell} \, \xi_{\epsilon} =
\epsilon \, \xi_{\epsilon}$. However, as $u$ is a Gamow vector,
$\xi_{\epsilon}$ diverges at the origin as $r^{-\ell -1}$. In other words,
$\xi_{\epsilon}$ is out of ${\cal H}$ and it is deprived of a physical
meaning. The same situation arises by considering the two--dimensional
kernel of the product $BA$. Hence, there are no more square--integrable
solutions of $h_{\ell}$ and the discrete spectrum $\sigma_d(h_{\ell})$ is
just the same as $\sigma_d(H_{\ell})$.

We have then constructed a non--hermitian Hamiltonian $h_{\ell}$ which is
strictly isospectral to the initial spherically symmetric Hermitian
Hamiltonian $H_{\ell}$. A simple calculation shows that the global
behaviour of the new potential $v_{\ell}(r) = V_{\ell}(r) - 2 u'(r)/u(r)$
is as follows
\be
v_{\ell}(r)= \left\{
\begin{array}{lr}
V_{\ell +1} (r) & r \sim 0\\[1ex]
0 & r \rightarrow \infty.
\end{array}
\right.
\label{pot}
\ee
Thus, for small distances, a particle with energy $E_n(\ell)$ interacts
with the field as having a quantum number $\ell +1$. In the asymptotic
region the particle behaves as free of interaction. On the other hand, the
intermediate region could be interpreted as `opaque' in the sense that the
particle interacts with a series of wells and barriers which alternate
their positions in the real and imaginary parts of $v_{\ell}(r)$ (see the
discussion on the optical bench given in \cite{Mie04}). The next section
elucidates the applications of the method by transforming the Coulomb
potential.

\subsection{Non-hermitian Hamiltonians with hydrogen--like
spectrum}

If the radial potential in (\ref{hamil1}) is the Coulomb
one $V(r) = -2/r$, the convenient Gamow vectors are given by the
expression (see Figure~\ref{potgam1}):
\be
u(r)= r^{\ell +1} e^{-kr} {}_1F_1 (\ell +1 - 1/k, 2\ell +2, 2kr)
\label{gamow}
\ee
with ${}_1F_1(a,c,z)$ the Kummer's function. The units of energy
and coordinates are respectively taken as ${\cal E} = Ze^2/2r_B$
(= $Z$ 13.5 ev) and $r_B = \hbar^2/Ze^2m$ (= $0.529 \times
10^{-8}/Z$ cm). The solutions (\ref{gamow}) have been explicitly
derived in \cite{Ros03}.

}

\begin{figure}[htbp]
\centering \epsfig{file=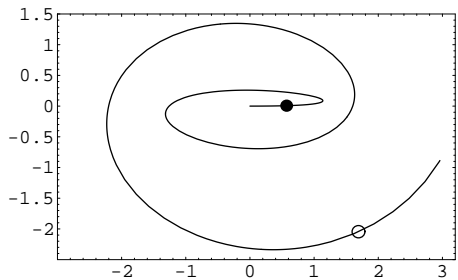, width=6cm}
\caption{\footnotesize The Argand-Wessel diagram of the hydrogen's Gamow
vector $u(r)$, from zero to 20 Bohr radii $r_B$, with $\ell=1$ and
$\epsilon = -0.2604 + i \, 0.104 \, (\times 13.5 \rm {ev})$. Horizontal
scale stands for the real part. The disk is at $r=1 \, r_B$ and the circle
at $r=19 \, r_B$.}
\label{potgam1}
\end{figure} 

\begin{figure}[htbp]
\centering \epsfig{file=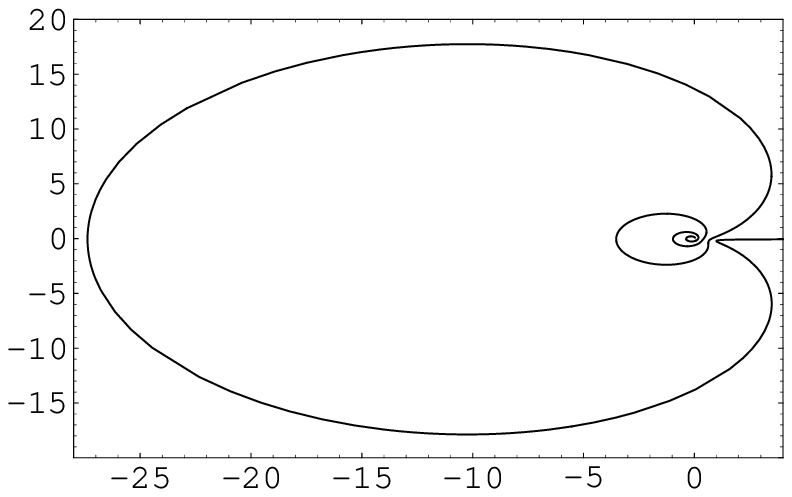, width=5cm}
\hspace*{1cm} \epsfig{file=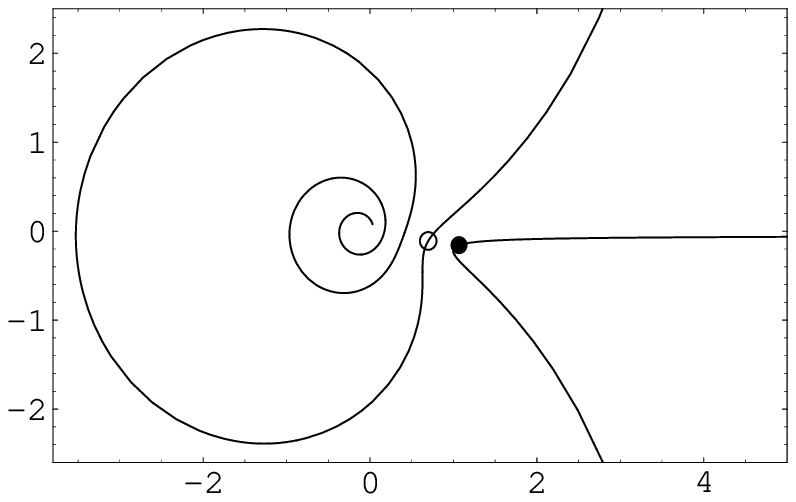, width=5cm}
\caption{\footnotesize The Argand-Wessel diagram of the non--hermitian
potential (\ref{pot}) for the same parameters as in Figure~\ref{potgam1}.
This potential has a discrete spectrum identical with that of the hydrogen
atom. The right--hand side figure is a detail of the cardiod--type one.
The disk is at $r=2 \, r_B$ and the circle at $r=6 \, r_B$.}
\label{cardiod}
\end{figure}

Once these Gamow vectors have been used as transformation functions in
(\ref{factor2}), the non--hermitian potential $v_{\ell}(r)$ resembles a
cardiod curve as depicted in the complex plane (see Figure~\ref{cardiod}).
Notice that $v_{\ell}(r)$ becomes almost real for small distances and goes
to $+\infty$ on the real branch for $r=0$. On the other hand, this
potential goes to zero when $r \rightarrow +\infty$. The imaginary part of
$v_{\ell}(r)$ becomes relevant for intermediate distances (i.e. at
distances which are between 2 and 6 Bohr radii, for the parameters
considered in the figure).

\begin{figure}[htbp]
\centering \epsfig{file=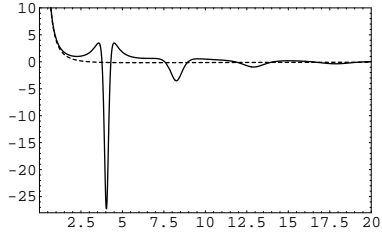, width=5cm}
\caption{\footnotesize The real part (continuous curve) of the
cardiod--type potential in Figure~\ref{cardiod} contrasted with the
effective Coulomb potential $V_{\ell =2}(r)$.}
\label{real1}
\end{figure}

Finally, Figure~\ref{real1} depicts the potential $V_{\ell +1}(r)$ as well
as the real part of $v_{\ell}(r)$. Observe the presence of barriers and
wells in the intermediate distances. A similar situation occurs for the
imaginary part of $v_{\ell}(r)$. These `partial potentials' induce local
`resonance' effects which are not present in the Hermitian potential
$V_{\ell +1}(r)$. Thus, the spatial distribution of the wave-packets
corresponding to $v_{\ell}(r)$ differ from that of the wave-packets of
$V_{\ell +1}(r)$ at the same energy.

\section{Generalized Gamow transformations}

As it has been shown in the precedent section, though the Gamow vectors
$u(r)$ could have a definite physical meaning as resonant states of
$H_{\ell}$, we consider them merely as transformation functions to
construct the non--hermitian Hamiltonians $h_{\ell}$. In general, all the
unphysical (not square--integrable) solutions of the Schr\"odinger
equation are useful to construct new Hamiltonians admitting real spectra
and square--integrable eigenfunctions \cite{nuestros,nuestros2}. In
particular, if the factorization constant $\epsilon$ is a real number, the
conventional factorization operators $A = B^{\dagger}$ are automatically
recovered.

Now, we extend the previous results by opening the chance to incorporate
complex eigenvalues with square--integrable solutions in the spectra of
the transformed Hamiltonians. First, notice that the general solution of
(\ref{eigen}) is, for small distances, a linear combination of two
particular solutions: $r^{\ell +1}$ and $r^{-\ell}$. The second one is
usually rejected because it is singular. Moreover, in the context of alpha
decay, a vector $u(r=0) \neq 0$ does not describe the `creation' of alpha
particles. We shall relax the Gamow condition at the origin to include the
solution $r^{-\ell}$ but preserving the purely outgoing condition
$e^{-kr}$. Let us remark that a `generalized' Gamow vector $\omega(r)$,
satisfying these new conditions, still is unphysical in the sense that it
is not square--integrable in ${\cal H}$.

The relevance of our generalization lies on the fact that expressions
(\ref{eigen})--(\ref{newsol}) still hold if $\omega(r)$ is taken as the
transformation function. Equation (\ref{cota2}), on the other hand, is
slightly modified:
\be
\Psi(r<<1) \sim \varphi'(r<<1) + \frac{\ell}{r} \, \varphi(r<<1).
\label{cota3}
\ee
Hence, the same conclusion is obtained: if $\varphi \in L^2(\mathbb R)$
then $\Psi \in L^2(\mathbb R)$. However, for complex eigenvalues of
$h_{\ell}(r)$, the kernel of $A$ provides the eigensolution
$\Psi_{\epsilon} \propto 1/\omega$, which can be normalized in ${\cal H}$
and satisfies $h_{\ell} \, \Psi_{\epsilon} = \epsilon \, \Psi_{\epsilon}$.
It is easy to check that the corresponding complex conjugate $\bar
\Psi_{\epsilon}$ is neither in the kernel of $A$ nor that of $BA$ (see
equation (\ref{factor2})). In counterdistinction, if $\varphi(r)$ is
eigensolution of $H_{\ell}$ belonging to $\epsilon$, then $\bar \varphi
(r)$ belongs to $\bar \epsilon$ as $H_{\ell}$ is selfadjoint.

Therefore, the discrete spectrum of $h_{\ell}$ is now given by
$\sigma_d(H_{\ell}) \cup \{ \epsilon\}$. On the other hand, the new
potential $v_{\ell}(r)= V_{\ell}(r) -2\omega'(r)/\omega(r)$ behaves in
this case as
\be
v_{\ell}(r)= \left\{
\begin{array}{lr}
V_{\ell -1} (r) & r \sim 0\\[1ex]
0 & r \rightarrow \infty
\end{array}
\right.
\label{pot2}
\ee
with a similar interpretation as for (\ref{pot}) but changing
$\ell +1$ by $\ell -1$. 

\begin{figure}[htbp]
\centering \epsfig{file=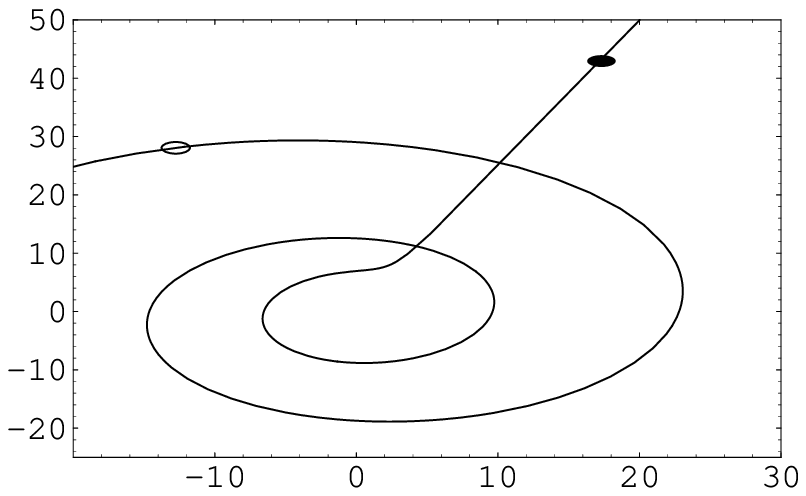, width=5cm}
\hspace*{1cm} \epsfig{file=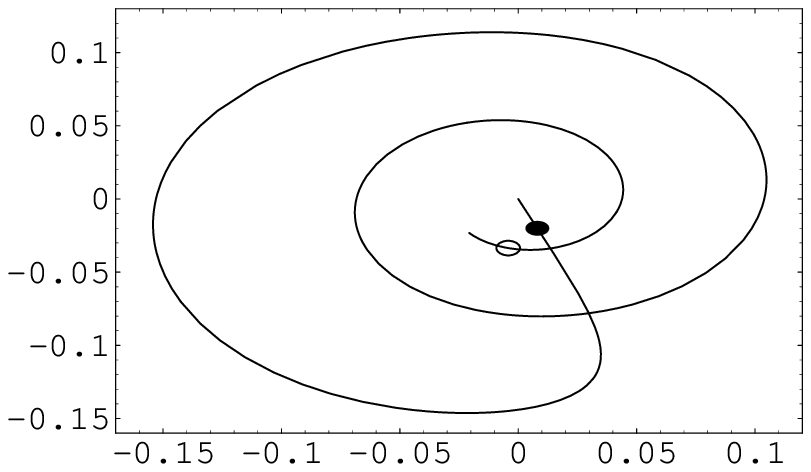, width=5cm}\\
{\bf (a)} \hspace*{5cm} {\bf (b)}
\caption{\footnotesize (a) The generalized hydrogen's Gamow vector
$\omega(r)$, plotted from zero to 20 Bohr radii and the same parameters as
in Figure~\ref{potgam1}. The disk is at $r= 0.05 \, r_B$ and the circle at
$r=19.5 \, r_B$. (b) The square--integrable wavefunction $\Psi_{\epsilon}$
belonging to the complex eigenvalue $\epsilon = -0.2604 + i\, 0.104 \,
(\times 13.5 {\rm ev})$. The disk is at $r= 0.05 \, r_B$ and the circle at
$r= 19 \, r_B$.}
\label{gral1}
\end{figure} 

\begin{figure}[htbp]
\centering \epsfig{file=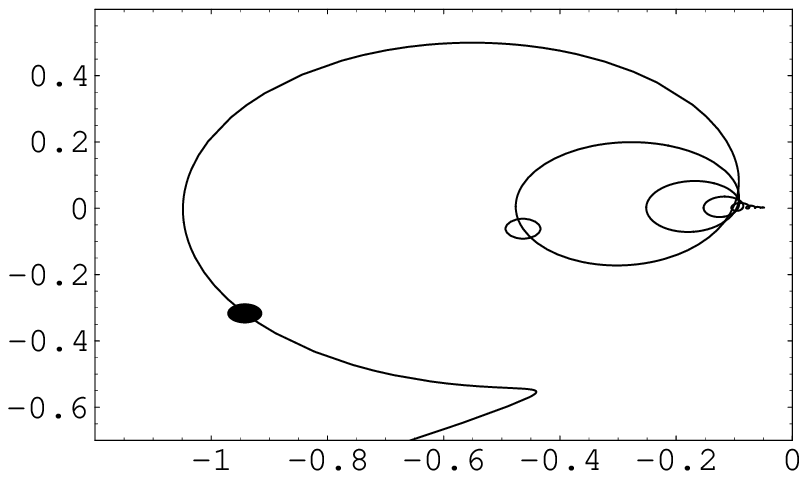, width=5cm}
\hspace*{1cm} \epsfig{file=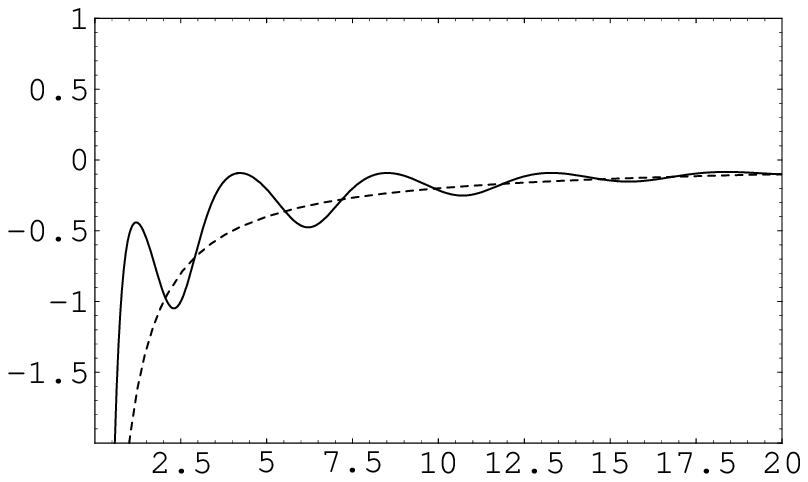, width=5cm}\\
{\bf (a)} \hspace*{5cm} {\bf (b)}
\caption{\footnotesize (a) The non--hermitian potential (\ref{pot2})
constructed via the generalized Gamow vector of Figure~\ref{gral1}a. Here
we have plotted from zero to 40 Bohr radii. The disk is at $r=2 \, r_B$
and the circle at $r= 6 \, r_B$. This potential has the same spectrum as
the hydrogen atom extended by an extra complex eigenvalue at $\epsilon$
with the square--integrable wavefunction of Figure~\ref{gral1}b. (b) The
corresponding real part (continuous curve) contrasted with the Coulomb
potential $V_{\ell =0}(r)$.}
\label{gral2}
\end{figure}

Figures~\ref{gral1} and \ref{gral2} show respectively the behaviour of the
generalized Gamow vector $\omega(r)$, the wavefunction
$\Psi_{\epsilon}(r)$ and the new non--hermitian potential $v_{\ell}(r)$
for the Coulomb case $V(r)=-2/r$. The related transformation function
is \cite{Ros03}:
\be
\begin{array}{rcl}
\omega(r) &=& r^{\ell +1} e^{-kr}[ {}_1F_1 (\ell +1 - 1/k, 2\ell
+2,
2kr)\\[1ex]
& & + \,\,\,  \xi \, U(\ell +1 - 1/k, 2\ell +2, 2kr)]
\end{array}
\label{gamow2}
\ee
where $\xi$ is a complex constant and $U(a,c,z)$ is the logarithmic
hypergeometric function.

\section{Concluding remarks}

The Gamow (decaying) eigensolutions have been shown to be appropriate
transformation functions in the framework of supersymmetric quantum
mechanics. Non--hermitian Hamiltonians, which are supersymmetric partners
of spherically symmetric self--adjoint energy operators, have been
constructed so that they admit purely real spectrum with normalized
wavefunctions. For other of these new non--hermitian operators an extra
complex eigenvalue with square--integrable eigensolution is present.

At first sight, our results could be connected with those derived in the
complex--scaling method. However, the Gamow--transformed potential
$v_{\ell}(r)=V_{\ell}(r) - 2 \omega'(r)/\omega(r)$ is not as simple as the
complex--scaled potential $V_{\theta} = e^{i\, 2 \theta} \,
V_{\ell}(re^{i\, \theta})$. In general, an intertwined Hamiltonian
$hB=BH$, $HA=Ah$ (factorized in a refined way \cite{refined}: $H=AB +
\epsilon$, $h= BA + \epsilon$) could correspond to a complex--scaled
Hamiltonian $h_{\theta} S =SH$ if $h_{\theta}=h$. Thus, there must exist a
couple of differential operators $M=AS$ and $N=BS^{-1}$, such that
$[H,M]=[h,N]=0$. A particular case has been recently reported \cite{Fer00}
(see also \cite{Neg00}) by considering conventional factorization
operators $a^{\pm} = \mp \frac{d}{dr} + \alpha(r)$, $\alpha : \mathbb R
\mapsto \mathbb R$, real factorization constants ${\cal E}$, and the
squeezing operator $S=U_r = e^{i\, \frac{\lambda}{2} \{ r, p\}}$,
$[r,p]=i$. The so derived `scaled intertwined' Hamiltonian $h_{\lambda}$
has a real potential $v_{\lambda}(r) = e^{2\lambda} \, V(e^{\lambda} r) -
\alpha'(r)$ and real discrete spectrum $\sigma_d(h_{\lambda}) = \{ e^{2
\lambda} {\cal E}, e^{2 \lambda} E_n\}_{n \in \mathbb N}$, where $e^{2
\lambda} {\cal E}$ is the ground state energy and $E_n \in \sigma_d(H)$.
This procedure allows to deform the excited energy levels of $h_{\lambda}$
but leaving unaffected the ground state ${\cal E}_0$:
$\sigma_d(h_{\lambda}) \mapsto \{ {\cal E}_0, ({\cal E}_0/{\cal E}) E_n
\}$. This is a remarkable profile of the factorization which is rarely
considered in the literature. Thus, it seems that complex--scaled
Hamiltonians $h_{\theta}$ could be successfully constructed as an
application of the technique reported in \cite{Fer00}. Work in this
direction is in progress.

On the other hand, the possible connection of our results with other
approaches as the $PT$--symmetry \cite{Ben03} or the pseudo--hermitian
transformation \cite{Mos03} has been discussed in a previous work
\cite{Ros03}.

Finally, we have presented the case of first order, not mutually adjoint,
intertwining operators $A$, $B$. However, the method can be iterated at
will by considering $h_{\ell}$ as the new initial Hamiltonian. The $n$th
iterated result can be also obtained by means of intertwining operators of
$n$th order. In particular, the second order case can be properly used to
obtain self--adjoint Hamiltonians with spectrum identical to the initial
one \cite{Ros03,Fer03}. It is also possible to show that second order
transformations can produce non--hermitian operators with real spectrum
extended by two extra complex eigenvalues $\epsilon$ and $\bar \epsilon$.
These results will be published elsewhere.

\subsection*{Acknowledgements}

The author is grateful to the organizers of the V Taller de la DGFM-SMF,
Morelia Michoac\'an, Mexico, for the kind invitation. Special thanks are
due to U. Nucamendi for the warm hospitality. The author is indebted to M.
Lomeli for the manuscript's typing. The support of CONACyT is
acknowledged.


\end{document}